I SENT ON SATURDAY JULY 13TH. PLEASE DISREGARD IT. I AOPOLOGIZE 
\documentstyle[preprint,aps]{revtex} 
\begin{document}
\tightenlines 
\title
{Supersolid phase in fully frustrated Josephson junction arrays} 
  
\author{Luigi Amico, Giuseppe Falci, Rosario Fazio and Gaetano Giaquinta}
\address{
 Istituto di Fisica, Facolt\'a di Ingegneria, Universit\'a di Catania, 
viale A. Doria 6, I-95129 Catania, Italy }

\maketitle 
\begin{abstract}
We study the phase diagram and the excitation spectra of an array of 
small Josephson junctions at $f=1/2$ and arbitrary charge 
frustration. We find that the supersolid region in the phase diagram 
is larger than the correspondent region at $f=0$ and it includes 
two different phases.In the chiral supersolid (SS) charges and vortices 
are arranged in a checkerboard pattern on a $2\:\times \:2$ supercell 
analogously to the unfrustrated case. We find a new phase, which 
we term {\em non-chiral supersolid} (NCSS), which has no corresponding
phase at $f=0$.
The excitation spectra in the supersolid regions 
show  particle like dispersions which is related to the
defectons. The defecton  condensation leads to superfluidity in the presence 
of charge ordered background. 
\end{abstract}


\section{Introduction}

Josephson  Junctions Arrays (JJA) are ideal model systems to study 
a variety of phase transitions induced by thermal or 
quantum fluctuations~\cite{NATO,FRASCATI}. These latter 
play a major role if the superconducting islands are of submicron size and
drive the zero temperature Superconductor--Insulator (SI) phase 
transition.  
The two caracteristic energy scales in the system are the Josephson  energy 
$J$ which is associated to the  tunneling of Cooper pairs between 
neighboring islands and the charging energy $U$ which is the energy cost 
to add an extra Cooper pair on a neutral island. The electrostatic energy 
tends to inhibit the Josephson tunneling: indeed a finite $U$ produces
quantum fluctuations of the phases $\phi$'s of the superconducting order 
parameter  $\Delta \mbox{e}^{i \phi_i}$ one each island.
If $J \gg U$ the system is superconducting since the fluctuations of the 
$\phi$'s are small and the system is globally coherent. 
In the limiting case $J/U \rightarrow \infty$, which will be referred 
as the {\it {classical}} case, the electrostatic energy plays no role.
In the opposite limit, $J \ll U$, the 
array is a Mott insulator since strong quantum fluctuations of $\phi_i$
prevent the system to reach coherence (Coulomb blockade of Cooper pairs).  
The SI phase transition in JJA has  been studied in great detail 
both experimentally~\cite{SI-EXP} and theoretically~\cite{SI-TH}. 
The effect of disorder and the presence of an
additional glass transition has been studied in Ref.~\onlinecite{GLASS}. 
At the transition the system could be a metal with a universal value of the
conductance~\cite{FISHER,WEN,SI-CONDUCTANCE}.

Frustration in a quantum JJA can be introduced either by applying a magnetic
field or by means of a gate voltage with respect to the ground plane. The
effect of the {\em magnetic frustration} has been studied extensively 
in  the classical limit~\cite{FRUSTRATION}.
The presence of the magnetic field  induces vortices in the system and if
the frustration is a rational number ($f \equiv \Phi / \Phi_0 =p/q $ 
where $\Phi$
is the magnetic flux piercing each plaquette and $\Phi_0 = h/2e $ ) 
then the ground state consists of a checkerboard configuration of 
vortices on a $q\: \times \: q$ cell.
A particularly interesting case is the fully frustrated situation
($f=1/2$) where the two degenerate ground states consist of
vortex lattice with a 
$2 \: \times \: 2$ elementary supercell. The current corresponding to this 
vortex arrangement flows
either clockwise or anticlockwise in each plaquette. We refer to this
$\phi$'s configuration  as a  {\it chiral} (ground) state. 
The effect of quantum fluctuations for
JJA at $f=1/2$ has been investigated in Ref.~\onlinecite{FISHMAN}: 
besides driving the  SI transition at $T=0$, quantum fluctuations affect 
also the superconducting regime. Indeed they reduce both    
the superconducting transition temperature and   
the magnitude of Josephson currents (though the modulus  $\Delta$ 
of the order parameter is unchanged). However the configuration of the 
phases $\phi_i$ and the supercurrent flow patterns are unchanged 
so the ground state is still chiral.   

Quantum effects may be modulated by means of a gate voltage to the ground 
$V_x$, which we denote as {\em  charge frustration}. Indeed the 
energy difference for two charge states in each
island with $n$ and $n+1$ extra electrons may be reduced 
by changing $V_x$. Consequently
the effects of a finite charging energy are weakened and 
the superconducting region in the phase diagram turns out to be enlarged. 
In the presence of charge frustration and finite range Coulomb interaction, 
the phase diagram has a rather rich structure. For instance different Mott 
insulating phases are present. They are characterized by different 
crystal-like configuration of the extra charge  on each island $e n_i$
which is arranged in superlattices whose lattice constant depends
on $V_x$. In addition  new {\it supersolid } phases  may appear 
where off-diagonal (superfluid) and diagonal (charge-cristalline) long range order coexist.
Since the original prediction 
by Andreev and Lifshits~\cite{ANDREEV} there were numerous  works concerning
the determination of the location and the size of these supersolid regions
in the phase diagram~\cite{LIU,MATSUDA,BRUDER,STROUD}. 
Most of the theoretical investigations are restricted to the mean 
field approximation; a step beyond this, using Quantum Monte Carlo was done in 
Ref.~\onlinecite{OTTERLO} for the JJA Model and in
Ref.~\onlinecite{SCALETTAR} for the Bose-Hubbard model. The 
presence of the supersolid region is 
not an artifact of the mean field approximation though 
its size is substantially reduced 
due to the effect of fluctuations. Investigations in one dimensional
systems~\cite{NIYAZ} by means of Monte Carlo simulation do not show any
trace of supersolid phase.   

All these investigations concerning frustration in quantum arrays considered
either charge {\em or} magnetic frustration. 
The combined effect of {\it both} frustration may lead to
interesting effects. 
The most striking prediction is that for certain ratios
of the magnetic to charge frustration the JJA will be in a Quantum Hall
phase~\cite{QHE}. The proposal is rather suggesting although 
experiments are not yet avaliable to support it and there is not 
a detailed study of the phase diagram which would allow to located the 
region where the Quantum Hall phase should be observed.  

Motivated by these facts we make a first step in this direction 
by studying the phase diagram and the low lying excitations 
of a JJA in the presence of both charge and phase frustration. 
In particular we will confine ourselves to the case of fully frustrated JJA
for which the ground state is known in the classical limit.
Although we cannot make still statements concerning the QHE already at this
stage we find new interesting results. 
Two different supersolid phases are identified which we indicate as
{\em chiral} (SS) and {\em Non-Chiral} (NCSS) supersolid. This NCSS is a
new phase and has not a corresponding stable 
phase at $f=0$. 

The {\em whole} supersolid region is enlarged  as compared 
to the case of zero magnetic field.

The paper is organized as follows. In the next section we introduce the
Quantum Phase Model (QPM) commonly used to study JJA and reduce it 
in the standard way to the XXZ Heisemberg
model. 
The phase diagram of the latter model is studied at 
mean field level in section III using the $1/S$ expansion.
We would like to point out that the presence of both
magnetic and charge frustrations makes 
the calculation non trivial already at this level and that
it is impossible to use other standard tools as Monte Carlo simulations 
or the self-consistent harmonic approximation. In the case of
Monte Carlo there are sign problems while in the harmonic approximation 
the discreteness of charges which is essential to describe
supersolids is not accounted for. 
In section IV we study 
the spectrum of the low-lying excitations in all the regions of the phase 
diagram. Finally, section V is devoted to the  conclusions.
   
\section{The Model} 
  
The physics of JJA made of submicron size junctions 
is usually described in terms of
quantum dynamics of the phases $\phi_i$ (at low 
temperatures the fluctuations of the modulus $\Delta$ are unimportant).
Since in nanofabricated samples there are no  
Ohmic currents between the islands and also quasiparticle tunneling
is negligible all the relevant physics is captured by the 
QPM
\begin{equation}
H_{QP} =  \sum_{ {i} , {j} }  (n_{i}-n_x) \; U_{ij} \; (n_{j}-n_x) 
-J \sum_{< {i} , {j} >} 
\cos \left (  \phi_{i}-\phi_{j} - A_{ij} \right ) \; , 
\label{QPham}
\end{equation}
where $n_{i}$
and $\phi_{i}$ are canonically conjugated with,  
$[\phi_{i},n_{j}]=2e \,i \; 
\delta_{ij}$. 
The Coulomb interaction is described by the matrix 
$U_ {ij}= 4e^2C_{ij}^{-1}$ where $C_{ij}$ 
is the inverse of the capacitance matrix. 
The external voltage $V_x$ enters via the induced charge $en_x$ and 
fixes the average charge on each island.
A perpendicular magnetic field with vector potential ${\bf A}$ 
enters the QPM in the standard way through 
$A_{ij}= \frac{2 e}{\hbar c} \int_{i}^{j} {\bf A} {\bf \cdot} d {\bf l}$
and gives the magnetic frustration parameter 
$f= {1 \over 2 \pi} \; \sum A_{ij}$ where the summation runs on an 
elementary plaquette.
In the classical limit, $J / U \rightarrow \infty$, 
JJA's are 
a physical realization of the two dimensional XY model. 

We will consider very large {\em onsite} Coulomb interaction and
very low temperatures where few charge states are important. 
In this limit the relevant physics is captured by considering only two 
charge states per island and
the QPM is equivalent to a
spin-$1/2$ Heisenberg model\cite{LIU} 
\begin{eqnarray}
H_{S} = - h\sum_{i} \, S^{z}_{i} + \sum_{ {i} , {j} } 
 \, S^{z}_{i} \; U_{ij} \; S^{z}_{j} 
- J \sum_{\langle {i} , {j} \rangle}  
\left ( e^{iA_{ij}} \; S^{+}_{i} \; S^{-}_{j} +
e^{-iA_{ij}} \; S^{+}_{j} \; S^{-}_{i} \right ) \;,
\label{Sham}
\end{eqnarray}
where the operators $ S^{z}_{i}, \; S^{+}_{i},\;S^{-}_{j}  $ are the 
usual $su(2)$ operators, $ S^{z}_{i}$ being related to the 
charge on each island ($n_{i}=S^{z}_{i}+ \frac{1}{2}$),
and the raising and lowering $S^{\pm}_{i}$ operators 
corresponding to the "creation" and "annihilation" operators 
$e^{\pm i \phi_{j}}$ of the QPM.
The  "external" field $h$ is 
defined as $h=\left (n_x-1/2\right ) \sum_{i,j} U_{i,j}$. 
A particulary interesting case is when $n_x$ is close to a 
half integer and the two charge states are almost 
degenerete.
The various magnetic orderings in the XXZ hamiltonian correspond to the
different phases in the QPM: long range order in $\langle S^x \rangle$ and
$\langle S^y \rangle$ 
($\sqrt { {\langle S^x_i \rangle}^2+{\langle S^y \rangle}^2}\neq 0$) 
indicates superfluidity in the QPM, while long range order in 
$ \langle S^z  \rangle $   ($ \mid \langle S^z \rangle \mid \neq 0 $)
describes order in the charge configuration.

In order to treat the 
$f=1/2$ case we consider a square lattice 
and use the gauge ${\bf A} = H y \hat {\bf x}$. 
This gives $A_{ij}=\pi \; {y_i \over a} \; \mbox{sign} (x_j-x_i)$ 
where the coordinates of the neighbouring sites $i$ and $j$ appear
and $a$ is the lattice spacing.
This is illustrated in Fig.(\ref{plaquette}) where the dashed lines 
indicate
the antiferromagnetic bond ($A_{ij}= \pi$) 
and the continous lines the ferromagnetic bonds ($A_{ij}= 0$). 

In the classical limit there are two degenerate $2 \times 2$ periodic 
ground states: in the QPM language they correspond to two characteristic 
$\phi_i$ patterns (one of them is shown in Fig.~\ref{plaquette}) which 
determine a checkerboard arrangement of
vortices and antivortices.

In the $XXZ$ version this corresponds
to   $\langle S^x_i \rangle $ and $\langle S^y_i \rangle $
reproducing the above $\phi $ pattern,
$ \arctan \left ( \langle S^y_i \rangle  /  \langle S^x_i \rangle \right )
=\phi_i$ 
and and $\sqrt { {\langle S^x_i \rangle}^2+{\langle S^y_i \rangle}^2}=1$.
Charge are completely delocalized  which corresponds to 
$\langle S^z_i \rangle =0$.

In the rest of the paper we 
consider the Coulomb interaction only between nearest neighbour 
($U_1$) and next nearest neighbour  ($U_2$) sites.

\section{Phase Diagram}

In this section we obtain the phase diagram for $f=1/2$ in the presence 
of arbitrary charge frustration. Previous work already established 
the  phase diagram of the spin hamiltonian in zero magnetic 
field~\cite{LIU,MATSUDA,BRUDER,SCALETTAR}. 
The chiral configuration of the classical ground states suggests the 
introduction of four sublattices indicated by the index $l= \alpha ,\;
\beta ,\; \gamma \; \delta $ (Fig. \ref{plaquette}). Each site is 
parametrized by $l$ and  an additional index $p$ which runs within each 
sublattice.

We now focus on  the XXZ model which we study using the $1/S$-expansion 
(S is the modulus of the spin vector). 
First the quantization axis in each sublattice is rotated to align 
the spins along the direction of the relative magnetization. 
We obtain the rotated hamiltonian ${\bf R} H_{S} {\bf R}^{-1}$,
where
\begin{equation}
{\bf R}\doteq \prod _p \prod_{l= \alpha ,\;\beta ,\; \gamma \; \delta } 
{R^{z}}_{pl} {R^{x}}_{pl} \; ,
\end{equation}
with ${R^{z}}_{pl} \doteq  e^{ i\phi_{l} S^{z}_{pl} } \;,
{R^{x}}_{pl} \doteq  e^{ i\theta_{l} S^{x}_{pl} } $. 
Then we perform the $1/S$ expansion and determine 
the actual values of the eight angles $\phi_{l}$ 
and $\theta_{l}$ by minimizing the 
$S \rightarrow \infty$ limit $H_{\infty}$ of the rotated 
hamiltonian. 
The above procedure allows to characterize ground states whose 
properties are uniform within each sublattice. 
The order parameters will be $sin \theta_l$  
for global phase coherence and $cos \theta_l$
for charge ordering. The $\phi_l$ configuration will determine
the supercurrent configuration in the flux phases. 

The $1/S$ expansion of the rotated hamiltonian can be 
carried on sistematically by using the Holstein-Primakoff 
transformation 
$
	S^{+}_{pl} = \sqrt{2 S} \, a_{pl}^{\dagger} \, 
	\sqrt{1-n_{pl}/S}\; ,\: 
	S^{-}_{pl} = \left ( {S^{+}_{pl}}\right 
	)^{\dagger}\; ,\: 
	S^{z}_{pl} = n_{pl}-S \;\:,
$
where $a_{pl}\;,a_{pl}^{\dagger}$ are bosons, and 
$n_{pl} = a_{pl}^{\dagger} a_{pl}$ 
These operators describe excitations around the $S \rightarrow \infty$ 
ground state (and {\it not} particles in the QPM).
We obtain
\begin{equation}
{\bf R} H_{S} {\bf R}^{-1}=H_{\infty}+H_{SW} +
{\cal O} \left (\sqrt {S} \right ) \; .
\end{equation} 
 where $H_{\infty}$ is of order $S^2$,  $H_{SW}$ 
describes low energy fluctuations around the $S \rightarrow \infty$
ground state and turns out to be of order $S$.
The $S \rightarrow \infty$  ground state properties 
are obtained by minimizing $H_{\infty}$ which is
 a sum over $p$ of identical terms of the form  
\begin{eqnarray}
H_{\infty} &=& {N \over 4} \; \sum_l \left[
 -\frac{h}{2} \cos \theta_l+\frac{ U_{1} }{8}  
\cos \theta_l
\sum_{m=nn \left (l \right ) } 
\cos \theta_{m} +\frac{U_2}{4} \cos \theta_l
\sum_{m=nnn \left (l \right ) } \cos \theta_{m} \right.
\nonumber \\
&& \qquad \qquad - \left. \frac{J}{4} \sin {\theta_l }
\sum_{m=nn \left (l \right ) } e^{A_{lm}} \sin \theta_{m} 
\cos \left ( \phi_l-\phi_{m} \right ) \right]
\label{hinfinity}
\end{eqnarray} 
The minimization of the $H_{\infty}$  leads to eight equations in the 
variables  $\phi_l$ and $\theta_l$  defined on the plaquette 
of Fig.(\ref{plaquette}.b).
Four of these equations can be solved analitically with the result  

\begin{eqnarray}
	& \tan &\left ( \phi_{\gamma}-\phi_{\beta} \right ) = 
	\frac{\sqrt {16 a^{2} b^{2}- \left ( b^{2}-1 \right )
	\left ( a^{2}-1 \right )}}
	{\left ( b^{2}-1 \right ) \left ( a^{2}+1 \right )} \; ,\nonumber \\
	& \tan & \left (\phi_{\delta} -\phi_{\alpha}\right )= -
	\frac{\sqrt {16 a^{2} b^{2}- \left ( b^{2}-1 \right ) 
	\left ( a^{2}-1 \right )} }
	{\left ( a^{2}-1 \right ) \left ( b^{2}+1 \right )} \; ,\nonumber \\
	& \tan & \left (\phi_{\beta}-\phi_{\alpha}\right )= -\frac{\sqrt {16 a^{2} b^{2}- 
	\left ( b^{2}-1 \right ) \left ( a^{2}-1 \right )} }
	{\left ( a^{2}-1 \right ) \left ( b^{2}+1 \right )-2 b^{2} 
	\left (a^{2}+1 \right)} \;.
\label{phisol}
\end{eqnarray}
where
$$
a = \frac{\sin \theta_{\alpha}}{\sin \theta_{\delta}}, \; 
b = \frac{\sin \theta_{\beta}}{\sin \theta_{\gamma}}
$$

We now look for solutions with 
$a=b=1$ which correpond to phases with checkerboard simmetry. 
The resulting configuration of the $\phi_l$ is
the same as in the classical limit, so the supercurrent 
has the same chiral pattern.
the ground state is chiral as in the classical situation. This has been
noticed in the case of zero external charge in Ref~\onlinecite{FISHMAN}.
We  show that for non zero charge frustration 
the above conclusion holds true for the superfluid phase and for
the chiral SS supersolid. 

The remaining 
equations for $\theta_l$ reduce to the ones of the $f=0$ case with the
substitution $J \rightarrow \frac{\sqrt{2}}{2} J $; 
the solutions then read   
\begin{equation}
\cos \theta = \frac{h}{2 \left ( U_{1}+U_2+\sqrt {2} J \right )}.
\end{equation}
for the superfluid (SF) and 
\begin{equation}
	\begin{array}{c}
	\cos {\theta}_{\alpha} \\ 
	\cos {\theta}_{\beta}
	\end{array} 
	 =
	\left\{
	\begin{array}{c}
	\frac{1}{2} \left ( v + \sqrt {v^{2}-4 w} \right )\\
	\frac{1}{2} \left ( v + \sqrt {v^{2}-4 w} \right )
	\end{array}
	\;\right. ,
\end{equation}
the chiral SS supersolid. 
We defined 
$w \doteq  (h-h_{s})/\left (2 U_2 \kappa \right )\;$ ,$
h_{s}  \doteq  2 \sqrt { {\left ( U_{1}-U_2 \right )}^{2}-{ 2 J^2}} \;$,
$\kappa  \doteq  \sqrt { 
1-{ \left [\sqrt {2} J/\left ( U_{1}-U_2 \right ) \right ]}^{2}
} \left ( 1+w \right ) $.

By compairing the energies relative to the above solutions 
we obtain the boundaries between the phases with ceckerboard simmetry

\vspace{0.2cm}
\noindent 
\underline{Paramagnetic - canted state} 
\begin{equation}
 h=\pm 2 \left (U_{1}+U_2+\sqrt {2} J \right ) \; ,
\end{equation}
{\underline{Canted state - $\frac{1}{4}$ or   
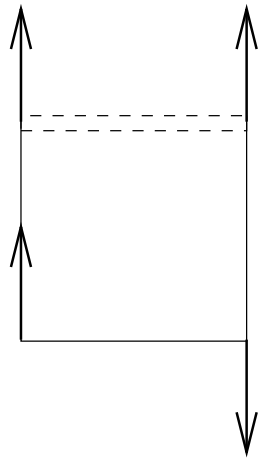
$\frac{3}{4}$ lobe} 
\begin{equation}
h=\pm \left [ U_{1}+U_2+\sqrt {2} J+\sqrt {  \left ( U_{1}+U_2+\sqrt {2} J \right )
\left ( U_{1}+U_2-3 \sqrt {2} J \right )}\right ] \; ,
\end{equation} 
\underline{Canted - chiral supersolid state}
\begin{equation}
h=2 \left ( U_{1}+U_2+\sqrt {2} J \right ) 
\sqrt {\frac{\left ( U_{1}-U_2-\sqrt {2} J \right )}{\left( U_{1}-U_2+\sqrt {2} J \right )}} \;,
\end{equation}
\underline{Chiral supersolid state - N\`eel state}
\begin{equation}
h=2 \sqrt { {\left ( U_{1}-U_2 \right )}^{2}-{ 2 J^2}} \; ,
\end{equation}
\underline{Chiral supersolid - $\frac{1}{4}$ or   
$\frac{3}{4}$ lobes}
\begin{equation}
h=\pm \left [ 2 U_2+h_{s}-\sqrt{
{\left (2 U_2+h_{s} \right )}^{2}-{h_{s}}^2-8 U_2 \left ( U_{1}-U_2 \right )
} \; \right ] \; .
\end{equation}

The spin configurations corresponding to the insulating, superfluid and
supersolid regions are shown in Figs.(~\ref{configurations}). In the Mott
lobes there is no projection of the spin on the $xy$ plane and the three spin
configurations in Fig.(~\ref{configurations}.a) correspond to filling
$1,1/2$ and $3/4$. In Fig.(~\ref{configurations}.b) and Fig.(~\ref{configurations}.c) 
the superfluid and supersolid regions are represented. 
The fact that the $\phi$ configuration is not affected by quantum 
fluctuations has already been noticed in Ref.~\onlinecite{FISHMAN}
for zero external charge. We extend this result to the SF and SS phases
resulting at non zero charge frustration. However we discuss below the NCSS
solution which does not show the above supercurrent pattern.

The NCSS ground state is found 
if we look for solutions with no ceckerboard 
simmetry. A non-ceckerboard phase was also found at $f=0$ by
Bruder {\it et al}~\cite{} and was named $SS2$ supersolid. 
We find two degenerate NCSS configurations namely
$\phi_{\alpha}=\phi_{\beta}=\phi_{\gamma}=0$ 
($\phi_{\delta}$ is meaningless since $sin \theta_{\delta}$ turns out
to vanish) and 
$\phi_{\alpha}=0$, $\phi_{\beta}=\phi_{\delta}=\pi$, 
($\sin \theta_{\gamma}=0$).   
We obtained the regions of the phase diagram where 
the NCSS solution is stable 
by minimizing numerically the energy (\ref{hinfinity}).
The resulting pattern of the supercurrent is different from 
the classical chiral one.
We point out that in this case the equations for the angles 
$\theta_l$ cannot be obtained from the $f=0$ equations by 
simply rescaling $J \rightarrow
J/\sqrt{2}$.

The solutions are characterized by the $sin \theta$ order parameter 
vanishing on one of the four sublattices. On this particular sublattice
$cos \theta = 1$ so the charge is well defined.
Thus the NCSS ground state describes a supersolid, since 
charge-crystalline and superfluid order coexist, but the chiral supercurrent
pattern is lost.
The phase boundaries obtained numerically and
the  configurations of the angles $\phi_l$ and $\theta_l$ of the NCSS ground
state are shown in Fig.(~\ref{configurations}.d). 

All the phase boundaries obtained in this section are summarized in
the phase diagram shown in Fig.(\ref{diagram1}). The {\em whole} 
supersolid region
(SS and NCSS phases) is enlarged compared to the $f=0$ case 
(SS1 and SS2 phases). At $f=0$ the tip of the lobe
(which coincides with the extension of the supersolid phase in the hard core 
limit) will correspond to $J/U_1 = 0.45$. 
A blow-up of the NCSS
region is shown in Fig.(\ref{diagram2}) and compared with the rescaled
$f=0$ phase diagram. 
This figure emphasizes that although the {\em whole} supersolid region 
is obtained by 
rescaling the $f=0$ phase diagram, the individual phases are not
(the rescaled
SS2-SS1 phase boundary is shown by crosses).

\section{Excitation Spectra}

The excitation spectra can be obtained calculating 
the eigenmodes of $H_{SW}$. As pointed out in the previous 
section all the terms up to ${\cal O} \left (S \right )$ 
are retained in the $1/S-$expansion; this, in turns, corresponds 
to  retain products at most bilinear  of the creation and annihilation 
operators. If we regarded  $l$ as a colour index,  
$H_{SW}$ describes a system of four  kinds of interacting $l$-bosons 
defined on a lattice whose sites are labeled by the index $p$.
Then we Fourier transform with respect to $p$ and   $H_{SW}$ reduces 
to a sum of single mode hamiltonians

\begin{eqnarray}
H_{SW}&=&\sum_{k} H_{k} \; , \nonumber \\
H_{k}&=& \left (
\varepsilon_{k}^{\left (\alpha, \beta \right )}+
\varepsilon_{k}^{\left (\alpha, \gamma \right )}+
\varepsilon_{k}^{\left (\alpha, \delta \right )}
\right ) n_{k,\alpha}+
\left ( 
\varepsilon_{k}^{\left (\alpha, \beta \right )}+
\varepsilon_{k}^{\left (\beta, \delta \right )}+
\varepsilon_{k}^{\left (\beta, \gamma \right )} 
\right ) n_{k,\beta} +  \nonumber \\
&&\left (
\varepsilon_{k}^{\left (\alpha, \gamma \right )}+
\varepsilon_{k}^{\left (\gamma, \delta \right )}+
\varepsilon_{k}^{\left (\beta, \gamma \right )} 
\right ) n_{k,\gamma}+ 
\left (
\varepsilon_{k}^{\left (\alpha, \delta \right )}+
\varepsilon_{k}^{\left (\beta, \delta \right )}+
\varepsilon_{k}^{\left (\gamma, \delta \right )}
\right ) n_{k,\delta}+  \nonumber \\
&&\left (v_{k}^{\left (\alpha ,\gamma \right )} 
a^{\dagger}_{k,\alpha} a_{k,\gamma}+
{v_{k}^{\left (\alpha ,\gamma \right )}}^{*} 
a^{\dagger}_{k,\gamma} a_{k,\alpha} \right )+
\left (v_{k}^{\left (\alpha ,\beta \right )} 
a^{\dagger}_{k,\alpha} a_{k,\beta}+
{v_{k}^{\left (\alpha ,\beta \right )}}^{*} 
a^{\dagger}_{k,\beta} a_{k,\alpha} \right )+ \nonumber \\
&&\left (v_{k}^{\left (\beta ,\delta \right )} 
a^{\dagger}_{k,\beta} a_{k,\delta}+
{v_{k}^{\left (\beta ,\delta \right )}}^{*} 
a^{\dagger}_{k,\delta} a_{k,\beta} \right )+
\left (  {v_{k}}^{\left ( \gamma ,\delta \right ) } 
a^{\dagger}_{k,\gamma} a_{k,\delta}+
{v_{k}^{\left (\gamma ,\delta \right )}}^{*} 
a^{\dagger}_{k,\delta} a_{k,\gamma} \right )+ \nonumber \\
&&\left (v_{k}^{\left (\alpha ,\delta \right )} 
a^{\dagger}_{k,\alpha} a_{k,\delta}+
{v_{k}^{\left (\alpha ,\delta \right )}}^{*} 
a^{\dagger}_{k,\alpha} a_{k,\delta} \right )+ 
\left (v_{k}^{\left (\beta ,\gamma \right )} 
a^{\dagger}_{k,\beta} a_{k,\gamma}+
{v_{k}^{\left (\beta ,\gamma \right )}}^{*} 
a^{\dagger}_{k,\gamma} a_{k,\beta} \right )+  \nonumber \\
&&\left (  q_{k}^{\left (\alpha ,\gamma \right )} 
a^{\dagger}_{k,\alpha} a^{\dagger}_{k,\gamma}+
{q_{k}^{\left (\alpha ,\gamma \right )}}^{*} 
a_{k,\gamma} a_{k,\alpha} \right )+
\left (q_{k}^{\left (\alpha ,\beta \right )} 
a^{\dagger}_{k,\alpha} a^{\dagger}_{k,\beta}+
{q_{k}^{\left (\alpha ,\beta \right )}}^{*} 
a_{k,\beta} a_{k,\alpha} \right )+ \nonumber \\
&&\left (q_{k}^{\left (\beta ,\delta \right )} 
a^{\dagger}_{k,\beta} a^{\dagger}_{k,\delta}+
{q_{k}^{\left (\beta ,\delta \right )}}^{*} 
a_{k,\beta} a_{k,\delta} \right )+
\left (q_{k}^{\left (\gamma ,\delta \right )} 
a^{\dagger}_{k,\gamma} a^{\dagger}_{k,\delta}+
{q_{k}^{\left (\gamma ,\delta \right )}}^{*} 
a_{k,\gamma} a_{k,\delta} \right )+ \nonumber \\
&&\left (q_{k}^{\left (\alpha ,\delta \right )} 
a^{\dagger}_{k,\alpha} a^{\dagger}_{k,\delta}+
{q_{k}^{\left (\alpha ,\delta \right )}}^{*} 
a_{k,\alpha} a_{k,\delta} \right )+  
\left (  q_{k}^{\left (\beta ,\gamma \right )} 
a^{\dagger}_{k,\beta} a^{\dagger}_{k,\gamma}+
{q_{k}^{\left (\beta ,\gamma \right )}}^{*} 
a_{k,\gamma} a_{k,\beta} \right ) \; ,
\label{HSW}
\end{eqnarray}

The coefficients in the hamiltonian (\ref{HSW}) (reported in the 
Appendix A)  depend on the angles $\theta_{l}$ and $\phi_{l}$ introduced
in the section III.
The $k$-sum is restricted to half of the Brillouin zone because of the 
doubling of the lattice constant due to the magnetic frustration. 
Linear contributions in the operators  $ a^{\dagger}_{k,i}\; , 
a_{k,i}$ vanish for the  $\{ \theta_l \;,\; \phi_l \}$ ground state 
configuration.

The hamiltonian $H_{SW}$ is diagonalized using an algebraic 
technique~\cite{WYB,MONTORSI} briefly described below:
we introduce the following operators 
\begin{eqnarray}
X_{ll^{\prime}} &\doteq& a_{k,l} a_{-k,l^{\prime}}+a_{k,l^{\prime}} a_{-k,l} \; ,\nonumber \\
X^{ll^{\prime}} &\doteq& a_{k,l^{\prime}}^{\dagger} a_{-k,l}^{\dagger}+
a_{k,l}^{\dagger} a_{-k,l^{\prime}}^{\dagger}=\left ( {X_{ll^{\prime}}}\right )^{\dagger} \; ,\nonumber \\
{X_{l}}^{l^{\prime}} &\doteq& a_{k,l^{\prime}}^{\dagger} a_{k,l}+
a_{-k,l^{\prime}}^{\dagger} a_{-k,l} \; ,\nonumber \\
{X_{l^{\prime}}}^{l} &\doteq& a_{k,l}^{\dagger} a_{k,l^{\prime}}+
a_{-k,l}^{\dagger} a_{-k,l^{\prime}} =\left ({{X_{l}}^{l^{\prime}}}\right )^{\dagger} \; ,\nonumber \\
{X_{l}}^{l}&\doteq& n_{k,l}+n_{-k,l}+1 \;,
\label{GEN}
\end{eqnarray}
which obey the  following commutation rules
\begin{eqnarray}
\lbrack  X_{ll^{\prime}},X_{mm^{\prime}}\rbrack&=& \lbrack X^{ll^{\prime}},
X^{mm^{\prime}} \rbrack=0 \nonumber \\
\lbrack  X_{ll^{\prime}},X^{mm^{\prime}}\rbrack&=& {X_{l}}^{m}{\delta_{l^{\prime}}}^{m^{\prime}}+
{X_{l}}^{m^{\prime}}{\delta_{l^{\prime}}}^{m}+{X_{l^{\prime}}}^{m}{\delta_{l}}^{m^{\prime}}+
{X_{l^{\prime}}}^{m^{\prime}}{\delta_{l}}^{m} \; , \nonumber \\
\lbrack  X_{ll^{\prime}},{X_{m}}^{m^{\prime}}\rbrack&=& X_{lm} {\delta_{l^{\prime}}}^{m^{\prime}}+
X_{l^{\prime}m} {\delta_{l}}^{m^{\prime}} \; , \nonumber \\
\lbrack  X^{ll^{\prime}},{X_{m}}^{m^{\prime}}\rbrack&=& - X^{lm^{\prime}} {\delta_{m}}^{l^{\prime}}-
X^{l^{\prime}m^{\prime}} {\delta_{m}}^{l} \; , \nonumber \\ 
\lbrack  {X_{l}}^{l^{\prime}},{X_{m}}^{m^{\prime}}\rbrack &=&  {X_{m}}^{l^{\prime}} {\delta_{l}}^{m^{\prime}}-
{X^{l}}^{m^{\prime}} {\delta_{m}}^{l^{\prime}} \; ,
\label{COMM}
\end{eqnarray}
where we have omitted the $k$-index in the $X$'s and   
$\{l,l^{\prime},m,m^{\prime} \}$ are colour  indeces. 
The 36 operators 
$\{X_{ll^{\prime}},X^{ll^{\prime}},{X_{l}}^{l^{\prime}},{X_{l^{\prime}}}^{l}, 
{X_{l}}^{l}\}$ form a base for  the non-compact symplectic Lie algebra 
$sp(8)_{k} $\cite{PERELOMOV}. The diagonal operators ${X_{l}}^{l}$ generate 
the  Cartan sublagebra of $sp(8)_{k} $ whose dimension is  
$4$ (equal to the rank of the algebra). The other off-diagonal operators, 
in number of 32, are  the  non-Cartan generators  of the algebra.
By using the definitions of Eq.(\ref{GEN}),  
one can see that $H_{k}$  belongs to $sp(8)_{k}$ since it can be  written 
as linear combination of  a subset of its  generators
\begin{eqnarray}
H_{k}&=& \sum_{l} D_{ll}^{(k)}\; {X_{l}}^{l}
+\sum_{l\neq l^{\prime}} {D_{ll^{\prime}}}^{(k)}\; {X_{l}}^
{l^{\prime}}
+\sum_{l \neq l^{\prime}} {R_{ll^{\prime}}}^{(k)}\; X_{ll^{\prime}} \;,
\label{SP8H}
\end{eqnarray}   
where ${R_{ll^{\prime}}}^{(k)}={{R_{l^{\prime}l}}^{(k)}}^{*}$ and 
${D_{ll^{\prime}}}^{(k)}={{D_{l^{\prime}l}}^{(k)}}^{*}$. 
(The coefficients in the expression 
(\ref{SP8H}) 
are listed in the Appendix B). Eq.(\ref{SP8H}) 
suggests that $H_{SW}$ posseses ${\cal A}={\bigoplus}_{k} {sp(8)}_{k}  $ as 
dynamical algebra, therefore we can diagonalize 
the Hamiltonian  using the  fundamental  
faithful ( {\it i.e.} preserving the commutation rules) Irreducible 
Representation (IRR) of ${\cal A}$. 
It is worthwhile to notice  that because of the non compacteness 
of $sp(8)$, every finite dimensional IRR of such algebra is not 
hermitian~\cite{GIL}. 
Despite  of this fact we can use IRR of ${\cal A}$ in order to obtain 
the correct eigenvalues of $H_{SW}$ (see Appendix A). 
Such eigenvalues  still depend on
$ \theta_l $ and $ \phi_l $:
the actual spectra in the various phases are  worked out 
by specifying the correspondent $S \rightarrow \infty$ ground 
state $\{ \theta_l \;\,\; \phi_l \}$ configuration for each phase.
The main results of this procedure are described below for the various
regions of the phase diagram.

{\underline {Insulating phases}}: \hspace{0.2cm}
Since the superfluid order parameter is zero there is no effect due to 
magnetic frustration. The Ne\`el and insulating $\frac{1}{4}$ (and 
$\frac{3}{4}$) phases are charge-modulated solids. The lowest lying
excitations are gapped and particle like. They correspond to 
particle - hole excitations. In Fig.~(~\ref{excinsulator}.a)
and Fig.~(~\ref{excinsulator}.b) we show the spectrum for the $1/2$ and
$3/4$ lobes.    
The Ne\`el solid is charge-modulated along both $x$ and $y$ directions
and the gapped branches are shown in Fig.~(~\ref{excinsulator}.a).
The four branches in the $3/4$ lobe come from the more complicated 
structure of the elementary cell. 
In this state  the transverse phonon like excitation, 
characteristic of  the diagonal 
long range order along rows and/or columns (where the charge is uniform) 
In addition we find   the particle like spectrum (with positive curvature at small $k$) 
which reflects charge modulation. This is shown in 
Fig.~(~\ref{excinsulator}.b) where the 
the dashed branch represents the phonon like excitation.

The Mott phases are not modulated in any direction, then  
the low lying excitations are phonon like . 

{\underline {SF phase}}:\hspace{0.2cm}
Because of the 
doubling of the elementary cell  of the lattice due to the magnetic 
field,  a gapped  branch is present  
in addition to a gapless mode which  linearly depend on $k$ at small $k$, 
see Fig.~(~\ref{excsuperfluid}). This was already discussed in the 
classical case by Ariosa {\em et al.}~\cite{FRUSTRATION}).
Long  range diagonal order existing along 
any lattice direction induces   
transverse phonon like excitations.

{\underline {Supersolid $SS$ phase}}:\hspace{0.2cm}
In the $SS$ phase both  diagonal and off-diagonal order modulation are 
present. 
In this state  the in-phase density fluctuations are coupled to the 
off-diagonal fluctuations and decreases 
the sound velocity because it reduces the superfluid density.
The supersolid is characterized also by a modulation of the charge. 
Such modulation of the number of particles  can be considered as 
defects of the lattice and their oscillations can be regarded as a  
fluctuation of the diagonal order on each 
site~\cite{CHENG}. The localized quasi-particle associated to 
the  collective oscillations of the defects   
were termed  as {\it defectons}~\cite{ANDREEV}.
The gapped branch reveals the particle like excitations consistently
with this physical picture.  
In Fig.(~\ref{excsupersolid}) the acustic branch is reminiscent of the 
superfluid order while the gapped branch with positive curvature is
the excitation spectrum for the defectons.

{\underline {NCSS phase}}:\hspace{0.2cm}
As in the previous case, the off-diagonal 
order is coupled with the density waves, but in this state there are 
four branches. The gapless one, reflects the superfluid 
nature of  this state. Since the mean field superfluid 
order parameter vanishes in a lattice site of the $2\times 2$ plaquette 
the sound velocity is strongly decreased.
Two of the three gapped curves take into account
the defects of the lattice, the other one means that in the system there
are  transverse phase-phonon excitations too. This is summarized in
in Fig.(~\ref{excncss}). 

An important issue which can be addressed by studying the excitation 
spectrum is the determination of the dynamical critical exponent at 
the various phase boundaries. It turns out that the fully frustration
does not change the critical exponents at the phase boundaries.
At the $SS$-Ne\`el solid transition the gap of the lower branch  of the 
solid vanishes as $k^2$ giving a critical exponent $z=2$ \cite{SCALETTAR}. 
At the SF-$SS$ supersolid transition the critical mode is 
$k=(\pi,\pi)$ and the roton minimum is at this  wave vector. 
At the phase boundary the  roton gap disappears linearly in $k$
giving $z=1$.

\section{Conclusions}
In this paper we considered the properties of a fully frustrated quantum 
Josephson Junction array in the presence of arbitrary charge frustration. 
We determined the mean field phase diagram at $T=0$ as well as the low 
lying excitation spectrum using a $1/S$ expansion for the equivalent XXZ 
model. 

At $f=1/2$ two kinds of supersolid phases, in addition to
the ordinary superfluid and insulating phases, are present. Besides 
the chiral supersolid SS which has an analog in the absence of magnetic
frustration we find the new NCSS supersolid phase which is non chiral.
In the case of the superfluid and the SS supersolid the main effect of
quantum fluctuation and charge frustration is to lower in magnitude 
the superfluid order parameter. The supercurrent pattern induced by the external 
magnetic field is however unaffected. In our mean field analysis this is
reflected
in a rescaling of $J$. In other words the magnetic
frustration fixes the current distribution while the charge frustration 
is responsible for the reduction of the magnitude of the 
superfluid order parameter. 
In the non chiral supersolid NCSS diagonal and 
off-diagonal long range order coexist in a non trivial way and the 
combined effect of charge and magnetic frustration cannot be separated
as in the other phases. 
The NCSS ground state has no checkerboard
simmetry since  the superfluid order parameter vanishes 
in one of the four sites 
of each plaquette and two flux quanta may be accomodated in four 
neighboring plaquettes. There is non chiral symmetry in this phase.
The NCSS phase exists only in the fully frustrated case and has 
no analog at $f=0$.
The {\it whole} (SS plus NCSS) supersolid region
in the phase diagram is enlarged compared to the unfrustrated case:
the supersolid state  better adjust to
the  periodicity  (on a $2 \times 2$ plaquettes elementary cell) 
induced by magnetic frustration than the SF phase and  
it is more favoured than in the $f=0$ case.
This may be important for the experiments because
the combination of charge and  magnetic frustration may 
help in detecting the supersolid phase. 

We also determined the low lying excitation spectrum of the system.
Due to the combined presence of magnetic and charge frustration the 
excitation spectra become more structured. They can be revealed, for
instance, by studying the anomalies in the $I-V$ characteristics
when Andreev current is injected into the array is coupled from 
a normal metal electrode~\cite{FALCI}. 

We are currently investigating different rational values of the magnetic 
frustration in order to study the possibility of a Quantum Hall phase 
as predicted in Ref.(\onlinecite{QHE}).

\acknowledgments 
We would like to thank C. Bruder, A. van Otterlo, K.H. Wagenblast, 
M. Rasetti, G. Sch\"on, G.T. Zimanyi for useful discussions.
The financial support of the European Community 
(HCM-network CHRX-CT93-0136) and of INFM (Italy) is gratefully
acknowledged. Two of us (R.F and  G.F.) aknowldge the warm hospitality of the 
Institut f\"ur Theoretische Festk\"orperphysik (Karlsruhe) where part of
the work was performed.

\appendix

\section{}

In this Appendix we list the coefficients of the hamiltonian (\ref{HSW})

\begin{eqnarray}
\varepsilon_{k}^{\left (l, m \right )} &\doteq & 
2 t e^{i A_{lm}}  
\sin \theta_{l} \sin \theta_{m} \cos z_{l,m}-
\cos \theta_{l} \cos \theta_{m} \; ,  \nonumber  \\
{\varepsilon}_{k}^{\left (l, n \right )} & \doteq & 
-2 U_2 \cos {\theta}_{l} \cos {\theta}_{m} \; .
\end{eqnarray}
The coefficients of the off diagonal operators are
\begin{eqnarray}
v_{k}^{\left (l ,m \right )}&\doteq& 
\cos \left ({\bf k}\cdot {\bf a}_{l,m}\right ) \{  t e^{i A_{lm}} 
\left [ \left (\cos \theta_{l} \cos \theta_{m} +
1 \right ) 
\cos z_{l, m} - 
i \left ( \cos  \theta_{l} +\cos \theta_{m} \right ) 
\sin z_{l, m} \right ] + \nonumber \\
&&\frac{1}{2} \sin \theta_{l} \sin \theta_{m} \} \; , \nonumber\\
v_{k}^{\left (l ,n \right )} &\doteq& 
U_2 \cos k_{y} \cos k_{x} \sin \theta_{l} \sin \theta_{m} \; \nonumber \\
q_{k}^{\left (l ,m \right )}&\doteq& 
\cos \left ({\bf k}\cdot {\bf a}_{l,m}\right )  \{ t e^{i A_{lm}} \left [ \left 
( \cos \theta_{l} \cos \theta_{m} -1
\right )
\cos z_{l, m} + 
i \left ( \cos  \theta_{l} +\cos \theta_{m} \right ) 
\sin z_{l, m} \right ] + \nonumber \\
&&\frac{1}{2} \sin \theta_{l} \sin \theta_{m} \} \; ,\nonumber \\
q_{k}^{\left (l ,m \right )}&\doteq& 
-U_2 \cos k_{y} \cos k_{x} \sin \theta_{l} \sin \theta_{m}  \;
\label{OFFDIAG} 
\end{eqnarray}

where $l$ and $n$ indicate {\it {n.n.}} and {\it {n.n.n.}} sites respectively.
In the preceding expressions  the lattice 
spacing vector ${\bf a}_{l,m}$  has both normalized components and   
$z_{l, m}$ are the superfluid order parameter's out of phase 
$\left ( \phi_{l} -\phi_{m}\right )$ on nearest neibourgh sites.
In the present case the coefficents (\ref{OFFDIAG})
are ${\bf C}$-numbers and they reduce to real ones  in the 
zero magnetic field case only.

\section{}

The IRR faithful representation of $ sp(8)$ we used consists in mapping the 
hamiltonian $H_{SW}$ in the matrix ${\cal M}\left (H_{SW}\right )$. 
Since that  $ sp(8)$ is a non compact algebra ${\cal M}$ turns tu be not
hermitian and it  has the following structure (we use the same notation 
as in  Ref.~\onlinecite{ZHANG})

\begin{equation}
\left(
\begin{array}{cc}
{\cal D} & {\cal R} \\
-{\cal R} & -\tilde{{\cal D}} 
\end{array}
\right)
\end{equation}
where the tilde indicates the reflection in the minor diagonal 
($\tilde{{\cal R}} = {\cal R}$) and ${\cal D}$ is hermitian. The matrix 
elements of ${\cal D}$ are 
    
\begin{eqnarray}
{D_{11}}^{(k)}&\doteq & \frac{1}{2}\left ( 
\varepsilon_{k}^{\left (\alpha, \gamma \right )} +
\varepsilon_{k}^{\left (\alpha, \beta \right )}+
\varepsilon_{k}^{\left (\alpha, \delta \right )}+
h\cos \theta_{\alpha}\right )\; ,\nonumber \\  
{D_{22}}^{(k)}&\doteq & \frac{1}{2}\left ( 
\varepsilon_{k}^{\left (\alpha, \beta \right )} +
\varepsilon_{k}^{\left (\beta, \delta \right )}+
\varepsilon_{k}^{\left (\beta, \gamma \right )}+h\cos \theta_{\beta}
\right )\; ,\nonumber \\  
{D_{33}}^{(k)}&\doteq & \frac{1}{2}\left ( 
\varepsilon_{k}^{\left (\alpha, \beta \right )} +
\varepsilon_{k}^{\left (\beta, \delta \right )}+
\varepsilon_{k}^{\left (\beta, \gamma \right )}+h\cos \theta_{\beta}
\right )\; ,\nonumber \\  
{D_{44}}^{(k)}&\doteq & \frac{1}{2}\left ( 
\varepsilon_{k}^{\left (\gamma, \delta \right )} +
\varepsilon_{k}^{\left (\beta, \delta \right )}+
\varepsilon_{k}^{\left (\alpha, \delta \right )}+h\cos \theta_{\delta}
\right )\;. 
\end{eqnarray}

\begin{eqnarray}
{D_{12}}^{(k)}&\doteq& \frac{1}{2} v_{k}^{\left (\alpha ,\beta \right )}\; ,\;\;
{D_{13}}^{(k)}\doteq \frac{1}{2} v_{k}^{\left (\alpha ,\gamma \right )} \; ,\;\;
{D_{14}}^{(k)}\doteq \frac{1}{2} v_{k}^{\left (\alpha ,\delta \right )} \; ,\nonumber \\
{D_{23}}^{(k)}&\doteq& \frac{1}{2} v_{k}^{\left (\beta ,\gamma \right )} \; ,\;\;  
{D_{24}}^{(k)}\doteq \frac{1}{2} v_{k}^{\left (\beta ,\delta \right )}\; , \;\;  
{D_{34}}^{(k)}\doteq \frac{1}{2} v_{k}^{\left (\gamma ,\delta \right )} \;. 
\end{eqnarray}

The matrix ${\cal R}$ has the following structure
 
\begin{equation}
\left(
\begin{array}{cccc}
R_{14} & R_{13} & R_{12} & 0  \\
R_{24} & R_{23} & 0 & R_{21} \\
R_{34} & 0 & R_{32} & R_{31} \\
0 & R_{43} & R_{42} & R_{41} 
\end{array}
\right)
\end{equation}

and its elements are given below

\begin{eqnarray}
{R_{12}}^{(k)}&\doteq& \frac{1}{2} q_{k}^{\left (\alpha ,\beta \right )}\; ,\;\; 
{R_{13}}^{(k)}\doteq \frac{1}{2} q_{k}^{\left (\alpha ,\gamma \right )} \; ,\;\;
{R_{14}}^{(k)}\doteq \frac{1}{2} q_{k}^{\left (\alpha ,\delta \right )} \; ,\nonumber \\
{R_{23}}^{(k)}&\doteq& \frac{1}{2} q_{k}^{\left (\beta ,\gamma \right )} \; ,\;\;  
{R_{24}}^{(k)}\doteq \frac{1}{2} q_{k}^{\left (\beta ,\delta \right )}\; , \;\;  
{R_{34}}^{(k)}\doteq \frac{1}{2} q_{k}^{\left (\gamma ,\delta \right )} \;.
\end{eqnarray}

The  diagonalization 
of $H_{SW}$ is  equivalent  to an inner
automorphism of the algebra on itself.
In other words, we can define the unitary operator 
${\bf U}\doteq \prod_{k} U_{k}$, with
\begin{equation} 
U_{k}\doteq e^{G_{k}}\;,
\end{equation}
where the antihermitian  operator $G_{k}$ is a linear combination of the non-Cartan  
generators of the algebra  
\begin{equation}
G_{k} \doteq \sum_{r,s}  \left [ \psi_{r,s} \left (X_{rs}-X^{rs}\right )+
{\psi_{r}}^{s} \left ({X_{r}}^{s}-{X_{s}}^{r}\right ) \right ] \;.
\end{equation}
The rotation of the hamiltonian trough ${\bf U}$ defines an inner 
automorphism of the algebra generalizing the Bogoliubov transformation    
\begin{eqnarray}
{\bf U} H_{SW} {\bf U}^{-1}&=&\sum_{k} \exp \left (ad\; G_{k} \right ) H_{k} \;, \nonumber \\ 
 \exp \left (ad\; G_{k} \right ) H_{k}&\doteq& U_{k} H_{k} {U_{k}}^{-1}\nonumber \\
&=&H_{k}+\sum_{n} \frac{1}{n !}\
\lbrack ...\lbrack G_{k},\lbrack G_{k}, H_{k} \rbrack \rbrack ...\rbrack \;, 
\label{ROTA}
\end{eqnarray}    
For the closure property of the Lie 
algebras the equation (\ref{ROTA}) still  produce an element 
of ${\cal A}$, however  we can fix the $\psi_{rs}$'s and the ${\psi_{r}}^{s}$'s
in such a way to put to zero the coefficients of the off-diagonal part of the hamiltonian.
As  a final result $H_{SW}$ becomes proportional to the  
generators  of the Cartan subalgebra of ${\cal A}$   
\begin{equation}
{\bf U} H_{SW} {\bf U}^{-1}=H_{DIAG.}=\sum_{r=1}^{4} {\omega_{k}}^{(r)} {X_{r}}^{r}
\end{equation} 
In the  present case $H_{SW}$ should be diagonal after having solved 
16 coupled non linear equations. 
Being interested in the eigenvalues only, we are allowed to use the 
faithful representation of ${\cal A}$ to find the spectrum of the 
hamiltonian diagonalizing the matrix ${\cal M}\left ( H_{SW}\right )$ 
instead of $H_{SW}$ as operator using the equation (\ref{ROTA}). 
These operations are  equivalent  because the diagonalization procedure 
involves commutators between the elements of the algebra only.

\newpage
\begin{figure}
\caption{Due to the non trivial periodicity 
due to the magnetic frustration a $2 \times 2$ cell should be considered.
The labels for the sites as used in the paper are here reported a). 
Using periodicity boundary conditions, as shown in a), one can reduce the
problem defined on the plaquette drawn in b).}
\label{plaquette}
\end{figure}

\begin{figure}
\caption{ The 3-components spin vectors we used are characterized by 
the usual Euler angles $\theta$ and $\phi$. 
a)In the insulating phases the order parameter vanishes and only the 
$\theta$ configuration is important.
b)The mean field order parameter's phase configuration  is 
unaffected by magnetic frustration.
c) The $\theta$ and $\phi$ configurations
in  the supersolid $SS$. As in the superfluid state, the magnetic 
frustration preserves the chiral order of the ground state.
d)The two degenerate ground states of the NCSS.
The circle indicates the lattice site where
the phase of the superfluid order parameter is not defined.}
\label{configurations}
\end{figure}

\begin{figure}
\caption{The phase diagram for the fully frustrated JJA. 
The {\it n.n.n.} Coulomb interaction is  $U_{2}=0.1\; 
U_{1}$.}
\label{diagram1}
\end{figure}

\begin{figure}
\caption{ The region of the phase diagram containing the NCSS is shown
in detail. In addition we report (with crosses)
the phase boundary that should have the chiral supersolid 
$SS2$ by
rescaling $J$ as discussed in the text.}
\label{diagram2}
\end{figure}

\begin{figure}
\caption{a)The excitation spectrum of the  Ne\`el insulator.
at  $J=U_{1}/4$, $U_{2}=3U_{1}/8$ and $h=2U_{1}/3$. 
The behaviour at small $k$ reveals the particle like nature of 
the  excitations. In b) we show the four branches of the "3/4" 
insulating  phase ( $U_{2}=0.1U_{1}$, $h=2U_{1}$ and $J=0.1U_{1}$). 
The two lower curves have been  rescaled by a factor 10.}
\label{excinsulator}
\end{figure}

\begin{figure}
\caption{ The acustic and transverse phonon like branches in the superfluid  
are shown for $J=1 U_1,\; U_2=0.1\; U_1, \; h=2.2 \;U_1,\; 
k_y=0$. }
\label{excsuperfluid}
\end{figure}

\begin{figure}
\caption{ In this figure we show the dispersion relations in supersolid
$SS$ ($J=0.35 \, U_1$, $U_2=0.1 \, U_1$,  $h=1.7\, U_1$, $k_y=0$).
The sound velocity of the acustic branch is reduced compared to 
superfluid case. The gapped branche has a positive curvature at small $k$
characteristic of the particle like nature of the excitations.}
\label{excsupersolid}
\end{figure}

\begin{figure}
\caption{ The excitation spectrum of the NCSS phase ( 
$J=0.2 \, U_1$, $U_2=0.1 \, U_1$,  $h=1.848 \, U_1$, $k_y=0$). 
The two lower curves have been rescaled by a factor 5.}
\label{excncss}
\end{figure}

\begin{figure}
\caption{ At the phase boundary between the  Ne\`el insulator and 
Supersolid $SS$ the excitation spectrum vanishes as $k^2$. 
The transition is signaled by softening of the acustic branch 
proper of the supersolid ($J=0.1\, U_1$, $U_2=0.1\, U_1$, 
$k_y=0$).}
\label{neelssboundary}
\end{figure}

\begin{figure}
\caption{ At  Superfluid-Supersolid $SS$ phase transition the roton mode  
vanishes as $k$ ($J=0.5\; U_1,\; U_2=0.1\; U_1, \; k_x=k_y$).}
\label{supssboundary}
\end{figure}


\begin{references}
\bibitem{NATO} Proceedings of the NATO Advanced Research Workshop 
	on {\it Coherence in superconducting networks},  J.E. Mooji
	and G.Sh\"on Eds., Physica  {\bf B152} (1988)
\bibitem{FRASCATI} Proc. of the Conference on {\it Macroscopic quantum
	phenomena and coherence in superconducting networks} C. Giovannella
	and M. Tinkham Eds. , World Scientific (Singapore, 1995)
\bibitem{SI-EXP} L.J. Geerligs, M. Peters, L.E.M. de Groot,
	A. Verbruggen, and J.E. Mooij, Phys. Rev. Lett. {\bf
	63}, 326 (1989); H.S.J van der Zant, L.J. Geerligs, and
	J.E. Mooij, Europhys. Lett. {\bf 19}, 541 (1992);
	 C.D. Chen, P. Delsing, D.B. Haviland, Y.Harada, 
	and T. Claeson, Phys. Rev. {\bf B50}, 3959 (1995).
\bibitem{SI-TH}K.B. Efetov Sov. Phys. JETP {\bf 51}, 1015 (1980); 
	S. Doniach, Phys. Rev. B {\bf 24}, 5063 (1981); 
	R.Fazio and G. Sch\"{o}n, Phys. Rev. B {\bf 43}, 5307 (1991).
\bibitem{GLASS}M.P.A. Fisher, B.~P. Weichman, G. Grinstein, 
	and D.~S. Fisher, Phys. Rev. B {\bf 40}, 546 (1989).
	G.~G. Batrouni, R.~T. Scalettar, and G.~T. Zimanyi,
	Phys. Rev. Lett. {\bf 65}, 1765 (1990);
	W. Krauth and N. Trivedi,Europhys. Lett. {\bf 14}, 627 (1991).
\bibitem{FISHER} M.P.A. Fisher, G. Grinstein and S.M. Girvin, 
	Phys. Rev. Lett. {\bf 64}, (1990) 587.
\bibitem{WEN} X.G. Wen and A. Zee, Int.J.Mod.Phys. {\bf B4}, 437 (1990).
\bibitem{SI-CONDUCTANCE}M.-C. Cha, M.P.A. Fisher, S.M. Girvin, 
	M. Wallin, A.P. Young, Phys. Rev. B {\bf 44}, (1991) 6883;
	M. Wallin, E.S. S{\o}rensen, S.M. Girvin, and
	A.P. Young, Phys. Rev. {\bf B45}, 13136 (1992);
	G.G. Batrouni, B. Larson, R.T. Scalettar, 
	J. Tobochnik, and J. Wang, Phys. Rev. {\bf B48}, 9628  (1993);
	A. van Otterlo, K.-H. Wagenblast, R. Fazio, 
	and G. Sch\"on,  Phys. Rev. {\bf 48} (1993) 3316; 
	R. Fazio and D. Zappal\`a, Phys. Rev.{\bf B53}, R8883 (1996).
\bibitem{FRUSTRATION} S. Teitel and C. Jayaprakash, Phys. Rev. {\bf 27}
	(1983) 598; S. Teitel and C. Jayaprakash, Phys. Rev. Lett. {\bf 51}
	(1983) 1999; T.C. Halsey, Phys. Rev. {\bf B 31} (1985)5728;
	W. Y. Shih and D. Stroud, Phys. Rev. B {\bf 30} 6774, (1984); 
	W. Y. Shih and D. Stroud, Phys. Rev. B {\bf 28} 6575, (1983);
	D. Ariosa, A. Vallat and H. Beck., J. Phys. France {\bf 51}, 
	(1990) 
\bibitem{FISHMAN} Fishman, R. S. and Stroud, D., Phys. Rev. B {\bf 37} 
	1499, (1987)
\bibitem{LIU}K.S. Liu and M.E. Fisher, J. Low Temp. Phys. {\bf 10},
	655 (1973).
\bibitem{MATSUDA} H. Matsuda and T. Tsuneto, Suppl. Prog. Theor. Phys. 
	{\bf 46}, 411 (1970). 
\bibitem{BRUDER} C. Bruder, R. Fazio,  and G. Sh\"on, Phys. Rev. B {\bf 47}, 
	342 (1993).
\bibitem{STROUD} E. Roddick and D.H. Stroud, Phys. Rev. B {\bf 48} 16600 
	(1993).
\bibitem{OTTERLO} A. van Otterlo, K-H. Wagenblast, Phys. Rev. Lett. 
	{\bf 48}, 16600 (1994).
\bibitem{SCALETTAR} R.T. Scalettar, G.G. Batrouni, A.P. Kampf, G.T. Zimanyi
	Phys. Rev. B {\bf 51} 8467 (1995).
\bibitem{NIYAZ} P. Niyaz, R.T. Scalettar, C.Y. Fong and G.G. Batrouni,
	Phys. Rev. B {\bf 44} 7143 (1991) 
\bibitem{QHE} A. Stern Phys. Rev. B {\bf 50} 10092 (1994); 
	A.A. Odintsov and Yu. N. Nazarov Physica {\bf B203}, 513 (1994)
	and Phys. Rev. B {\bf 51} 1133 (1995);
	M.Y. Choi Phys. Rev. B {\bf 50} 10088 (1994).
\bibitem{CHENG} Y.C. Cheng, Phys. Rev.  B {\bf 23}, 157 (1981). 
\bibitem{CHESTER} G.Chester, Phys. Rev. A {\bf 2} 256, (1970).
\bibitem{ANDREEV} A.F. Andreev, I.M. Lifshitz, Sov.Phys. $JETP$ 
	{\bf 29}, 1107 (1969) 
\bibitem{PERELOMOV} Perelomov, A. M., 1986  {\it Generalized 
	Coherent States and their Applications}, (Springer, Berlin).
\bibitem{WYB} Wybourne B.G. 1974 {\it Classical  Groups  for Physicist}, 
	J.Wiley \& Sons.
\bibitem{SOLOMON} A.I. Solomon, J. Phys. A {\bf 14} (1981).
\bibitem{MONTORSI} A. Montorsi, M. Rasetti, A.I. Solomon, Phys. Rev. Lett.
	{\bf 59} 2243 (1987). 
\bibitem{AMICO} L. Amico, M. Rasetti, R. Zecchina, to appear in 
	Physica {\bf A}.
\bibitem{GIL} Gilmore R.,1974 {\it Lie Groups, Lie Algebra, 
	and some of Their Applications} J.Wiley \& Sons.
\bibitem{ZHANG} W.-M. Zhang, D.H. Feng and R. Gilmore, Rev. Mod. Phys. 
	{\bf 62}, 867 (1990).
\bibitem{FALCI} G. Falci, R. Fazio, A. Tagliacozzo, and G. Giaquinta,
	Europhys. Lett {\bf 30}, 169 (1995)

\end{references}
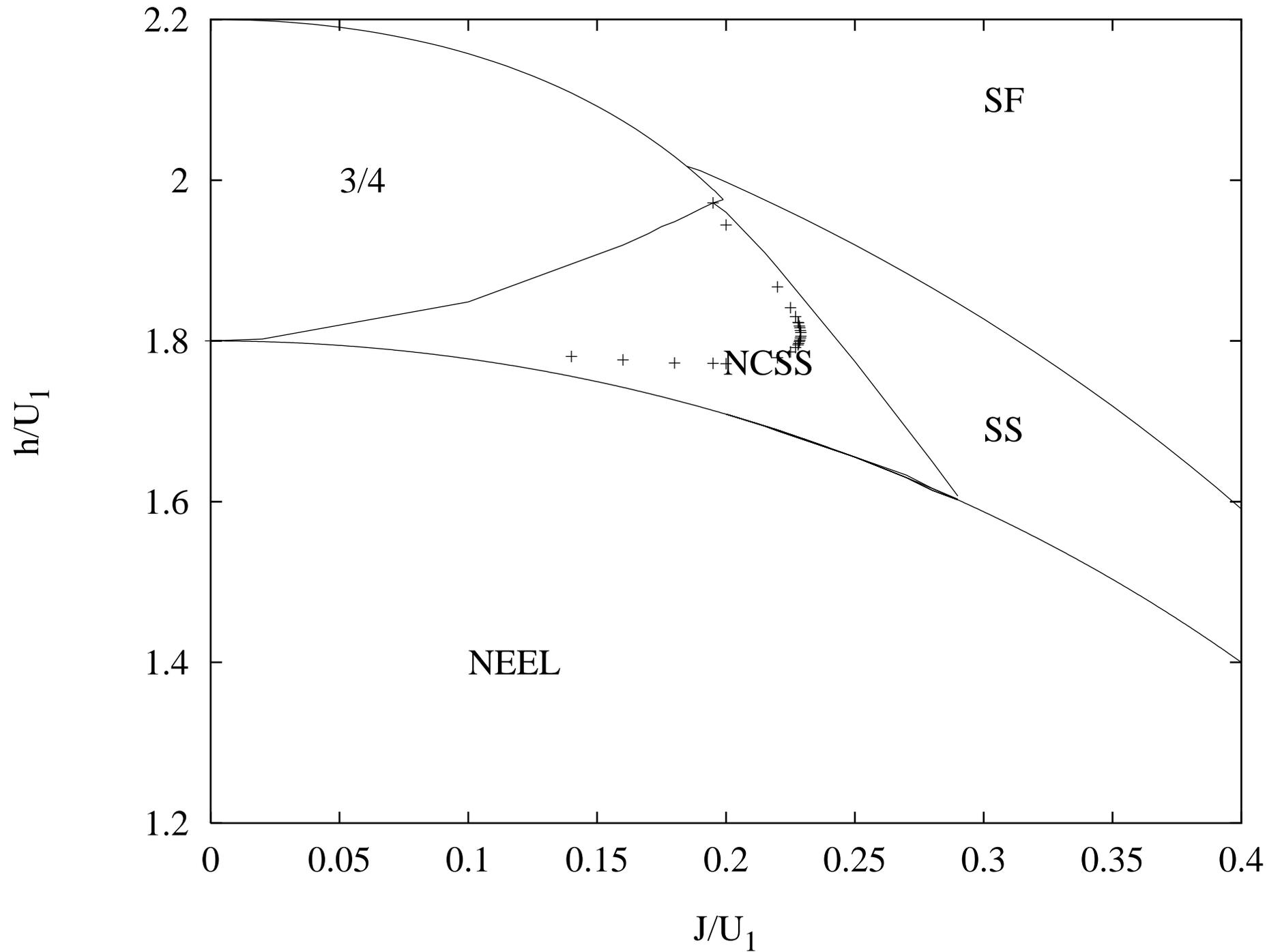
\end{document}